\documentclass[a4paper,11pt]{article}
\usepackage{latexsym}	
\usepackage{amsmath}
\usepackage{amsthm}

\usepackage{ amssymb }
\def\ba{\begin{eqnarray}}
\def\ea{\end{eqnarray}}

\def\ba{\begin{eqnarray}}
\def\ea{\end{eqnarray}}

\def\lb{\label}
\def\be{\begin{equation}}
\def\ee{\end{equation}}

\theoremstyle{plain}

\begin{document}

\baselineskip0.25in
\title{Some comments about  emission channels of non abelian vortices}
 \author{ Osvaldo P. Santill\'an \thanks{Instituto de Matem\'atica Luis Santal\'o (IMAS), UBA CONICET, Buenos Aires, Argentina
firenzecita@hotmail.com and osantil@dm.uba.ar.}}

\date {}
\maketitle

\begin{abstract}
As is well established, several gauge theories admit vortices whose mean life time is very large. In some cases, this stability is a consequence of the topology of the symmetry group of the underlying theory. The main focus of the present work is, given a putative vortex, to determine if it is non abelian or not by analysis of its physical effects. The example considered here is the simplest one namely, a $SU(2)$ gauge model whose internal orientational space is described by $S^2$. Axion and gravitational emission are mainly considered. It is found that the non abelian property is basically reflected in a deviation of gravitational loop factor $\gamma_l$ found in \cite{vachaspati}-\cite{burden}. The axion emission instead, is not very sensitive to non abelianity, at least for this simple model. Another important discrepancy is that no point of the vortex reaches the speed of light when orientational modes are excited. In addition, the total power corresponding to each of these channels is compared, thus adapting the results of \cite{davis}-\cite{peloso} to  the non abelian context. The excitations considered here are simple generalizations of rotating or spike string ansatz known in the literature \cite{ruso1}-\cite{kruczenski}. It is suggested however, that for certain type of semi-local strings whose internal moduli space is non compact, deviations due to non abelianity may be more pronounced.
 \end{abstract}

\section{Introduction}

The dynamic of gauge vortices is a fascinating branch of physics whose role is not yet well understood. Historically, it was realised by Abrikosov that magnetic field lines play a fundamental role in phase transitions in type low temperature II superconductors \cite{abrikosov}, and these objects were further studied in \cite{nielsen}. The dynamics of vortices in random environment is also of particular importance in the physics of high temperature superconductors \cite{anderson1}-\cite{anderson3}\footnote{See \cite{larkin} and its references for an extensive review.}. In addition, the phenomena of pinning of vortices may also have applications in the physics of neutron stars, as described for instance in \cite{sedrakian} and references therein. Abelian vortices were intensively studied as sources of galaxy formation, some classic works about this topic are \cite{vilenkin}-\cite{kibble}.

One notable prediction related to such vortices is the phenomena of linear confinement of magnets inside a low temperature type II superconductor \cite{abrikosov}. Based on these phenomena, Mandelstam, Nambu and 't Hooft suggested that a dual Meissner effect in which the field lines are chromo-electric, and  
the electric and magnetic charge are interchanged, may explain the long standing question about how quarks are confined inside the hadrons \cite{tof}-\cite{mandelstam}. The problem about this hypothesis is that it is not well understood how to include objects like monopoles in ordinary QCD, with  $SU(3)$ gauge group, in order to achieve this mechanism.

The line of work suggested in \cite{tof}-\cite{mandelstam}  took a great impulse with the work of Seiberg-Witten, in the context of supersymmetric theories \cite{seibergwitten1}-\cite{seibergwitten2}. These authors study the dynamics of light states of a  $SU(2)$ gauge theory with N$=2$ supersymmetry. This theory admits very massive monopoles, which becomes massless in certain limit of the parameter space. The model posses a duality that interchange the electric and magnetic field and charges
$$
\overline{E}\leftrightarrow \overline{B}, \qquad g\to \frac{1}{g},
$$
with $g$ the abelian coupling constant of the model. The addition of certain term that breaks N$=2$ supersymmetry to N$=1$ induce monopole condensation by Abrikosov lines. From the dual point of view, these lines are magnetic. But in the original theory, they are electric. This means that Seiberg and Witten found a realisation of a supersymmetric dual Meissner effect. 

One drawback of the Seiberg-Witten scenario is  that is related to abelian vortices, however it motivated a large amount of work about non abelian ones. In the context of supersymmetric theories, solutions of this type were found in \cite{navortex1}-\cite{navortex4}. Some of these models admit non abelian vortices when the s-quarks are the Higgs phase. These vortices induce monopole condensation at weak coupling, thus generalizing the Seiberg-Witten mechanism to the non abelian case. Another remarkable feature that arise in this context is the presence of phases  that are not identified neither with the Higgs, Coulomb or confined one \cite{seibergwitten3}. In particular, the "instead of confinement" phase considered in \cite{fradkinphase2}, in which the quarks and gauge bosons of the model decay into monopole and anti-monopole pairs that form stringy mesons \cite{fradkinphase}. This phase is continuously connected to the fully Higgsed phase. This bears a resemblance with the Fradkin-Shenker scenario \cite{fradkin} generalized to the supersymmetric context, but with the difference that the confined phase is replaced with the '"instead of confinement" one.

The vortices described above are related to N$=2$ supersymmetric gauge theories. Since their appearance, there have been investigations about these objects in theories with less supersymmetry \cite{esfuerzo1}-\cite{esfuerzo7}. In addition, some advances has been reported in the area of semilocal strings \cite{semilocal1} applied to supersymmetric theories \cite{semilocal2}-\cite{semilocal6}. In particular, an interesting link with critical superstrings was pointed out in \cite{novedad1} and further worked out in \cite{novedad2}-\cite{reviewshifman}. These works conjecture  that in the strong coupling regime, and in some specific thin limit, the resulting low energy theory can be identified with a IIA string over a target space which is the product of four dimensional space  with a six dimensional conifold. The Minkowski space represent the translational modes of the object, and the conifold represents internal modes of the vortex. At classical level, the conifold is not represented by a Ricci flat metric, but the conjecture takes into account quantum corrections. After these corrections have been properly taken into account it is believed that the Ricci flat  (Calabi-Yau) metric will emerge.  The string theory techniques then may be applied in order to study the spectrum of the states of the theory. More details can be found in \cite{novedad2}-\cite{reviewshifman}.

The physics of non abelian vortices is not only related to supersymmetric theories, and has in fact applications in ordinary QCD, even taking into account the drawback about monopoles mentioned above. In particular, the study of non abelian vortices has proven to be fruitful in the so called the colour-flavor locked phase of QCD \cite{cfl1}-\cite{cfl2}. This phase is supposed to appear for QCD at very high densities, such as the ones in the core of a neutron star. In this phase the mean distance between two hadrons is much less than its mean radius $r\sim$ fm, and is expected for the quarks composing these composite particles to acquire a large mobility. The relevant excitations in such high density phase are sourced by quarks close to the Fermi surface. These low energy excitations then have a very large momentum, which implies that the system is asymptotically free and the confined phase arguably does not take place  \cite{nitto2}. The gluons are now part of the asymptotic spectrum of the theory and induce an attractive interaction, giving rise to quark Cooper pairs which are not colour singlets. A very rough estimation of the resulting gap is  $\Delta\sim 50-100$ MeV, but there appear several corrections to this value due to the high chemical potential $\mu$ or the high temperature $T$ of the neutron star. The resulting state is symmetric under certain operation that interchange of color and flavour simultaneously, a colour-flavour diagonal symmetry \cite{lida}. For this reason this phase sometimes is referred as colour superconductivity or colour-flavor locked phase. Details of these affirmations may be found in the reviews \cite{nitto}-\cite{nitto2} and references therein. But is worthy to emphasize that the presence of a gap may affect the transport properties of this regions and may influence the cooling rates or their rotational properties of a neutron star \cite{shovkovy}. 
This phase, as well as other hypothetical QCD phases admits non abelian vortices, as reviewed in \cite{nitto}. Recent progress in the physics of these vortices have been reported in \cite{nitta1}-\cite{nitta24}. The colour-flavour diagonal symmetry, quotiented by a suitable subgroup, describe different inequivalent vortices.  Therefore these objects acquire a moduli, which is non abelian in nature. Details of these affirmations can be found in \cite{nitto} and references therein. 

The present work is focused on a simple and, at the moment, academic problem. This problem is, given an excited vortex, to understand if it is abelian or not abelian in nature by studying it emission channels. Particular attention is paid here on axion emission and also on gravitational waves. One of the main differences is that non abelian vortices may invest part of its energy in excitation of internal moduli. This in particular implies that there are no points in the vortex reaching the speed of light, as all the velocities are slowed for sourcing these internal excitations. Another important difference is the loop factor for gravitational radiation, whose value changes when internal orientations are excited, as will be discussed along the text. 

The organisation of this work is as follows. In section 2 some generalities about gauge theories admitting non abelian vortices are stated. In section 3 the dynamics of these vortices is characterised and some solutions are presented. In section 4, the coupling to axion particles is worked out and in section 5, formulas for the power radiated in axions are presented. The explicit power radiated for the presented solutions is estimated in section 6.  In section 7 the power radiated corresponding to gravitational wave emission is discussed and, in particular, it is clarified how non abelian excitations may affect it.   Section 8 also contains an axion radiation process, but in this case the internal modes excitations play a more important role than in the examples of section 6. Section 9 contains the discussion of the obtained results.

\section{Simple examples of non abelian vortices}

\subsection{N=$2$ supersymmetric gauge models}
In this subsection, some basic features about supersymmetric models and about the colour-flavor locked phase are briefly discussed, following \cite{navortex5} or \cite{nitto}. The  reader  familiar with these subjects may skip to the next subsection.

A typical (but not unique) form of a bosonic lagrangian for N$=2$ supersymmetric models, admitting non abelian vortices as solutions,
is the following \cite{navortex5}
$$
S=\int d^4x \bigg[\frac{1}{4g^2_2}F^{a}_{\mu\nu}F^{a\mu\nu} +
\frac{1}{4g^2_1}F_{\mu\nu}F^{\mu\nu}+
\frac{1}{g^2_2}|\nabla_{\mu}a^a|^2 +\frac{1}{g^2_1}
|\partial_{\mu}a|^2 
$$
\be\lb{boson}
+ {\rm Tr}|\nabla_{\mu}\Phi|^2 + {\rm Tr}|\nabla_{\mu} \bar{\tilde{\Phi}}|^2
+V(\Phi,\tilde{\Phi},a^a,a)\bigg]\,.
\ee
Here the gauge group is generically $SU(N)\times U(1)$ and $\nabla_{\mu}$ is the covariant derivative in the adjoint representation of the group
\be\lb{cov}
\nabla_\mu=\partial_\mu -\frac{i}{2} A_{\mu}
-i A^{a}_{\mu} T^a.
\ee
The coupling constants $g_1$ and $g_2$ correspond to the $U(1)$ and $SU(N)$  sectors respectively, and $a^a$ and $\Phi^{kA}$ are spin zero particles. The bosonic potential $V(\Phi^A,\tilde{\Phi}_A,a^a,a)$ is given by
$$
V(\Phi^A,\tilde{\Phi}_A,a^a,a) =
 \frac{g^2_2}{2}
\left( \frac{1}{g^2_2}\,  f^{abc} \bar a^b a^c+
 \bar{\Phi}_A\,T^a \Phi^A -\tilde{\Phi}_A T^a\,\bar{\tilde{\Phi}}^A\right)^2
+ \frac{g^2_1}{8}
\left(\bar{\Phi}_A \Phi^A - \tilde{\Phi}_A \bar{\tilde{\Phi}}^A -N \xi_3\right)^2
$$
$$
+\frac{1}{2}\sum_{A=1}^N \left\{ \left|(a+\sqrt{2}m_A +2T^a a^a)\Phi^A
\right|^2
 +\left|(a+\sqrt{2}m_A +2T^a a^a)\bar{\tilde{\Phi}}^A
\right|^2 \right\}
$$
\be\lb{pot}
+ 2g^2\left| \tilde{\Phi}_A T^a \Phi^A \right|^2+
\frac{g^2_1}{2}\left| \tilde{\Phi}_A \Phi^A -\frac{N}{2}\,\xi \right|^2,
\ee
with $f^{abc}$ the structure constants of the Lie algebra $SU(N)$.  The parameters $\xi_i$ come from Fayet-Illopoulos terms. In the following, the choice $\xi_3=0$, $\xi_2=0$ and  $\xi=\xi_1$ will be employed. By introducing the field
\be
A = \frac12\, a + T^a\, a^a,
\label{Phidef}
\ee
 the vacuum of the theory is parameterized as
\be
\langle A \rangle = - \frac1{\sqrt{2}}
 \left(
\begin{array}{ccc}
m_1 & \ldots & 0 \\
\ldots & \ldots & \ldots\\
0 & \ldots & m_N\\
\end{array}
\right).
\label{avev}
\ee
For generic values of the parameter $m_N$ the subgroup SU$(N)$ is broken to U(1)$^{N-1}$. 
However, for the specific choice of equal masses $m_1=m_2=...=m_N$, the classic group  SU$(N)\times$U(1) is not broken. The presence of the Fayet-Illopoulos parameter induce the following
non zero expectation values 
\be\lb{expect}\left<\Phi^{kA}\right>=\sqrt{\xi}\, \left(
\begin{array}{ccc}
1 & 0 & ...\\
... & ... & ... \\
... & 0 & 1  \\
\end{array}
\right),\qquad \left<\bar{\tilde{\Phi}}^{kA}\right>=0,
\ee
with $k=1,...,N,$ and  $A=1,...,N$.  The fact that the  squarks  $\Phi^{kA}$ and the gauge field $A$ acquire expectation values proportional to the identity matrix implies
that there exist the remanent symmetry 
$SU(N)_{C+F}$
\be\lb{cfl2}
\Phi\to U\Phi U^{-1},\qquad a^aT^a\to Ua^aT^aU^{-1},\qquad M\to U^{-1}MU,
\ee
with  $U$ an element of $SU(N)$, leaving invariant the vacuum of the theory. Such type of situations were already considered in the 70's in another context by Bardacki-Halpern \cite{bardacki}.
The symmetry (\ref{cfl2})  is of fundamental importance in the presence of vortices. The reason is that vortices break the diagonal symmetry, and different vortex solutions are then connected by a quotient $Q=SU(N)_{C+F}/ I$ of the action (\ref{cfl2}) with the group $I$ leaving these vortex solutions invariant. This give rise to internal moduli for these objects, described by this quotient $Q$.

A description similar to the one given above holds the colour-flavour locked phase \cite{cfl1}-\cite{cfl2}, with the Fayet-Illopoulos parameter $\xi$ replaced by the scale $\Delta_{cfl}\sim 50-100$ MeV. The details will not be made explict here, they can be found for instance in the review \cite{nitto}.

\subsection{The generic form of a simple non abelian vortex}
The models described above admit vortex solutions in general. In the following, the simplest type of non abelian type of vortices will be considered namely, vortices with moduli parameterized by the sphere $Q=S^2$. These vortices appear for instance in gauge scenarios with $SU(2)\times U(1)$ gauge group. The generic form for a vortex solution aligned along the $\hat{z}$ axis can be written as follows
$$
\Phi^{kA}=\Delta\; U\left(
\begin{array}{cc}
  e^{i\theta}\phi_1(r) & 0  \\
  0 &  \phi_2(r) \\
  \end{array}\right)U^{-1},
$$
\be\lb{beat}
A_{i}(x) = \frac{1}{2}\,\epsilon_{ij}\,\frac{x_j}{r^2}\,
[1-f_3(r)]U\tau^3 U^{-1},\qquad A_z=A_t=0.
\ee
Here the latin indices $i=1,2$ correspond to the $x$ and $y$ components of the $SU(2)$ gauge field $A_i$. The parameter $\Delta$ is  a characteristic  energy scale of the system.
It may be represent the gap $\Delta\sim 50-100$ MeV of the colour-flavour locked phase or the square root of the Fayet-Illopoulos parameter $\sqrt{\xi}$.
The scalars $\Phi^{kA}$ of the model compose a $2\times 2$ square matrix.   
The coordinates $r$ and $\theta$ are the standard polar coordinates on the plane defined by
$x=r\cos\theta$, $y=r\sin\theta$.
 The $SU(2)$ matrix $U$ is a global one, that is, it does not depend on the space time coordinates ($t$, $r$, $\theta$, $z$). 
 The following identity for these matrices
 \be\label{na}
U\tau^3 U^{-1}=n^a\tau^a,\qquad a=1,2,3,
\ee
is well known. The quantities $n^a$ represent a unit vector on $S^2$, that is, a vector satisfying $n^2=1$. The matrices $\tau_a$ are the standard Pauli matrices. The vortex solution can be expressed alternatively as
$$
\Phi^{kA}=\bigg[\frac{e^{i\theta}\phi_1(r)+\phi_2(r)}{2}\bigg] I+\bigg[\frac{e^{i\theta}\phi_1(r)-\phi_2(r)}{2}\bigg] n^a\tau^a,
$$
\be
A_{i}(x) = \frac{1}{2}\,\epsilon_{ij}\,\frac{x_j}{r^2}\,
[1-f_3(r)]n^a\tau^a,\qquad i=1,2.
\ee
Therefore, it is seen that the different vortices of the model are parameterized by the sphere $S^2$. This sphere of course, is not representing any geometry in the space $R^3$ or in the space-time $M_4$. Instead, it is a internal geometry describing different group elements characterizing all the possible vortex configurations.
The function $\phi_2(r)$ describing the scalar field in (\ref{beat}) is slowly varying. The function $\phi_1(r)$ instead is not, and it is zero in the $r=0$ line. In addition, $\phi_1(r)\to 1$ when $r\to\infty$.  For the supersymmetric case, the energy for unit length (tension) of the vortex \cite{navortex5}
 \be\label{eten}
T=2 \,\pi \, \xi,
\ee
 is independent on the chosen orientation $n^a$. For the colour-flavour locked phase, this tension is expected to be proportional to the symmetry breaking scale $\Delta_{cfl}$. It may roughly estimated as \cite{nitto}
\be\lb{tensito}
T\sim \frac{14 \zeta(3)}{72 \pi^3}\frac{\mu^2 \Delta_{cfl}^2}{T_c^2} \log L m.
\ee
Here $T_c$ is the QCD critical temperature, of the order $T_c\sim 100-150$ MeV and $\zeta(x)$ the Riemann zeta function. The mass $m$ is is related to an UV cutoff giving the vortex size core
$$
l_c\sim \frac{1}{m}\sim -\frac{96\pi^2 T_c^2}{7\zeta(3)}\log\frac{T}{T_c}.
$$
The cutoff $L$ is an IR one, and is related to large but finite dimensions of the system. The chemical potential $\mu$ is this phase is assumed to be very high, of the order of $\mu\sim 300$ MeV or even larger.

The vortex solution described above is static. The region where $\phi_1(r)$ vanishes is a line, which is interpreted as the string or vortex location.
One of the tasks of the present work is to study the decay channel of the vortex in axions. For this purpose, it is mandatory to describe the couplings between the axion $a$ and the gauge vector field $A_i$. The axion is a Goldstone boson and it is coupled to the vortex by an interaction term
\be\lb{axo}
S_a=\int_{M_4}\frac{a(x^\mu)}{f_a} \text{Tr}(F_{\mu\nu}\widetilde{F}^{\mu\nu})d^4x,
\ee
with $\widetilde{F}_{\mu\nu}$ the dual field strength corresponding to $F_{\mu\nu}$. The axion is not usually coupled directly to the gauge field, but 
this interaction is an effective one, induced by a triangle of heavy quarks in a ABJ Feymann diagram \cite{axion1}-\cite{axion4}.

\section{Vortex excitations}
\subsection{The excited vortex in the Manton regime}
Consider now a slightly excited vortex. The excitations arise by prompting the moduli $n^a$ of $S^2$ described in (\ref{na}) to a slowly varying field $n^a(z,t)$. Another type of excitation is obtained by deforming its shape. In this case the position of the vortex can fluctuate with a displacement $\delta x^\mu(z,t)$ around the static position $r=0$. 
For such excited vortex, the region of vanishing $\phi_1$ is a now a string $x^\mu(\upsilon^0,\upsilon^1)$ with time varying position in $R^3$. Here the coordinate $\upsilon^0$
is the temporal one while $\upsilon^1$ is the spatial one.
The coordinates $\upsilon^i$ swap a two dimensional surface, denoted  by $\Sigma$, which is interpreted as the worldsheet of the string. 
The equations of motions for the excited vortex, in the slow field or Manton approximation \cite{manton}, were obtained
in several references, see for instance \cite{navortex5} and references therein. In order to describe it, it is convenient to introduce six coordinates $s^\mu=(t,r,\theta, \phi, \alpha,\beta)$. The first four coordinates parameterize  the Minkowski space $M_4$ and describe the translation modes of the vortex. The last two coordinates describe the orientational $S^2$ field $n^a$ by the relation
\be\lb{nado}
n^1=\sin \alpha \sin \beta,\qquad n^2=\sin \alpha \cos\beta, \qquad n^3=\cos\alpha.
\ee
In these terms, the action describing the excitations of the vortex is \cite{navortex5}
\be\lb{manta}
S=T \int \sqrt{-|h|}h^{ab}g_{\mu\nu}\partial_a s^\mu \partial_b s^\nu d\tau d\sigma.
\ee
Here $g_{\mu\nu}$ is the canonical metric of $M_4\times S^2$
$$
g=-dt^2+dr^2+r^2(d\theta^2+\sin^2\theta d\phi^2)+R^2(d\alpha^2+\sin^2\alpha d\beta^2).
$$
The physical interpretation of the radius  $R$ of the orientational sphere $S^2$ deserve some comments. For the colour-flavour locked phase \cite{nitto}
it is given by $R^2\sim \mu^2 T^{-1}\Delta^{-2}_{cfl}$ with $T$ the vortex tension and $\mu$ the chemical potential of the environment where the vortices is located. It is estimated as
\be\lb{erre}
R\sim\frac{72 \pi^3}{14 \zeta(3)}\frac{T_c^2}{ \Delta_{cfl}^4\log L m} .
\ee
For the supersymmetric case instead, the radius is given by
\be\lb{erre2}
R\sim \frac{1}{\xi g_2^2},
\ee
with  the coupling $g_2$ defined in the lagrangian (\ref{boson}). The moral of these expressions is that, the larger the scale of broken symmetry is, the smaller the radius of the $S^2$ results. In addition, $h_{ab}$ denotes is the worldsheet metric of the string. It is an auxiliary field, as it does not contain any kinetic energy.

In order to solve the equations of motion arising from (\ref{manta}), it is customary, although not mandatory, to employ the conformal gauge $ \sqrt{-|h|}h^{ab}=\eta_{ab}=\text{diag}(-1,1)$. In the following the notation 
$\upsilon^0=\tau$, $\upsilon^1=\sigma$ will be employed. The lagrangian corresponding to (\ref{manta}) in the conformal gauge is then 
$$
{\cal L}=\partial_\tau t\;\partial_\tau t-\partial_\tau r\;\partial_\tau r-r^2\partial_\tau \theta\;\partial_\tau \theta-r^2\sin^2\theta\partial_\tau \phi\;\partial_\tau \phi-R^2\partial_\tau \alpha\;\partial_\tau \alpha-R^2\sin^2\alpha\partial_\tau \beta\;\partial_\tau \beta
$$
\be\lb{logrado}
-\partial_\sigma t\;\partial_\sigma t+\partial_\sigma r\;\partial_\sigma r+r^2\partial_\sigma \theta\;\partial_\sigma \theta+r^2\sin^2\theta\partial_\sigma \phi\;\partial_\sigma \phi+R^2\partial_\sigma \alpha\;\partial_\sigma \alpha+R^2\sin^2\alpha\partial_\sigma \beta\;\partial_\sigma \beta.
\ee
On the other hand, the two conformal constraints of the model are 
$$
-\partial_\tau t\;\partial_\tau t+\partial_\tau r\;\partial_\tau r+r^2\partial_\tau \theta\;\partial_\tau \theta+r^2\sin^2\theta\partial_\tau \phi\;\partial_\tau \phi+R^2\partial_\tau \alpha\;\partial_\tau \alpha+R^2\sin^2\alpha\partial_\tau \beta\;\partial_\tau \beta
$$
\be\lb{constro}
-\partial_\sigma t\;\partial_\sigma t+\partial_\sigma r\;\partial_\sigma r+r^2\partial_\sigma \theta\;\partial_\sigma \theta+r^2\sin^2\theta\partial_\sigma \phi\;\partial_\sigma \phi+R^2\partial_\sigma \alpha\;\partial_\sigma \alpha+R^2\sin^2\alpha\partial_\sigma \beta\;\partial_\sigma \beta=0,
\ee
and
\be\lb{constro2}
-\partial_\tau t\;\partial_\sigma t+\partial_\tau r\;\partial_\sigma r+r^2\partial_\tau \theta\;\partial_\sigma \theta+r^2\sin^2\theta\partial_\tau \phi\;\partial_\sigma \phi+R^2\partial_\tau \alpha\;\partial_\sigma \alpha+R^2\sin^2\alpha\partial_\tau \beta\;\partial_\sigma \beta=0.
\ee
The unperturbed string is given by $\tau=t$, $\sigma=z=r\cos\theta$, $\rho=r\sin\theta=0$, with $\alpha$ and $\beta$ fixed. This in particular implies that $\theta=0$ or $\theta=\pi$.
\subsection{Some simple excitations}
Consider now perturbed solutions. In the following the ansatz
\be\lb{roto}
\tau=t,\qquad \sigma=z, \qquad \phi=\omega t, \qquad \beta=\nu t,\qquad r=r(z),\qquad \theta=\alpha=\frac{\pi}{2},
\ee
will be considered. This ansatz bear some resemblance with classical string solutions considered for instance in \cite{ruso1}-\cite{ruso3}.
As now $\cos\theta=0$, this perturbation describes $z$ dependent oscillations in the radial cylindrical direction.
The second  conformal constraint is identically satisfied for this functional form of the excitation. Instead, the first one is not, and yields
$$
-1+\omega^2 r^2+R^2\nu^2+r'^2=0.
$$
Here the $'$ refers to a derivative with respect to $\sigma$. On the other hand, the equations of motion are simply
$$
r''+\omega^2r=0. 
$$
The last two equations are consistent, since the second arises by taking the derivative of the first with respect to $\sigma$. In addition, $R^2\nu^2<1$, which means that the limit frequency is $\nu=R^{-1}$.
The integration of the first equation throws the following result
\be\lb{roto2}
r=\frac{\sqrt{1-R^2\nu^2}}{\omega}|\sin \omega z|.
\ee
For an infinitely large string there is no constraint in $\omega$. 
In the following however, a large but finite string will be considered, with size $L$ much larger that its thickness and with fixed ends.
The presence of fixed endpoints require $r(z+L)=r(z)$, and this implies that $\omega=2\pi m/L$ with $m$ integer. At the end, it would be more
desirable to consider closed loops, as these are likely the main objects to appear in physical applications. However, in order to deal with the complications of non abelianity,
we will assume that these fixed end strings may approximate the excitations of a closed loop with radius $R=L/2\pi$.
The task is now to understand the energy loss of this object by axion emission.
\section{The coupling of the axion to the  orientational and translational modes of the vortex}
In order to study  the string energy loss by axion emission,  the couplings between the axion and the orientational and
translational modes of the vortex should be found. In order to figure out the orientational couplings, consider an excitation $n^a(z,t)$ of the unit vector defined in (\ref{na}).
The gauge field components $A_i$ are
assumed in this approximation to retain the same functional dependence on the coordinates, except that now $n^a\to n^a(z,t)$.
On the other hand, due to the non trivial dependence of $n_a$ with respect to $t$ and $z$, the gauge field components $A_\alpha$ with $\alpha=0,3$ are turned on.
A proper ansatz for  these components is  \cite{navortex5}
$$
A_\alpha=-i\rho(r)(\partial_\alpha U)U^{-1}.
$$
Note that, if the dependence on the moduli $n^a\to n^a(z,t)$ is neglected,
then this expression vanishes identically, and the solution (\ref{beat}) would be recovered. However, there are remanent $U(1)$
symmetries that leave the vortex solution invariant \cite{navortex5}. By a proper quotient
of this redundant action, it can be shown that  gauge field components may be written as \cite{navortex5}
\be\lb{newga}
2A_\alpha=-\rho(r)\epsilon^{abc}n^b\partial_\alpha n^c\tau^a,\qquad \alpha=0,3.
\ee
The field strength tensor is calculated by the convention
$$
F_{\mu\nu}=\partial_{\mu} A_{\nu}-\partial_{\nu} A_\mu-i[A_\mu, A_\nu].
$$
It is convenient
to write the components $i=1,2$ of (\ref{beat}) as
\be\lb{olga}
A_i=\epsilon_{ij} x^j g(r) n^a\tau^a,\qquad g(r)=\frac{1}{2r^2}[1-f_3(r)].
\ee
Then, by use of the convention just introduced, it can be calculated from (\ref{olga}) that
$$
F_{12}=F_{xy} =-[2 g(r)+g'(r)]n^a\tau^a.
$$
In addition, one has that
$$
F_{\alpha i}=\bigg[\frac{\partial_\alpha n^a}{r^2}\epsilon_{ij} x^j f_3(r)[1-\rho(r)]+\frac{x^i}{2r}\frac{d\rho}{dr}\epsilon^{abc}n^b \partial_\alpha n^c\bigg] \tau^a,
$$
$$
F_{30}=F_{zt}=-\rho(r)\epsilon^{abc}[\partial_{z} n^b\partial_{t} n^c-\partial_{t} n^b\partial_{z} n^c]\tau^a+\rho^2(r)\epsilon^{dbc}[\partial_{z} n^b\partial_{t} n^c-\partial_{t} n^b\partial_{z} n^c]n^d n^a\tau^a.
$$
In these terms, a simple calculation throws the following result
$$
\text{Tr}(F_{\mu\nu}\widetilde{F}^{\mu\nu})=h(r)\epsilon^{abc}[\partial_{z} n^b\partial_{t} n^c-\partial_{t} n^b\partial_{z} n^c]n^a,
$$
with $h(r)$ a function of $r$ whose explicit form is not very relevant for the following purposes, except that is arguably non vanishing for a region of the size of the vortex $\delta$. From this expression, by taking (\ref{axo}) into account,
the following induced coupling between the axion and the orientational modes
$$
S^{eff}_a=\alpha_a \int_{M_2}a(t,z,0,0)\epsilon^{abc}n^a \dot{n}^b n'^c dz dt,
$$
is obtained. Here $\alpha_a$ is a coupling arising due to the  integration over the transversal coordinates $x$ and $y$. It has dimensions $[\alpha_a]=$time=length in natural units. This effective action can be expressed alternatively as
\be\lb{ef}
S^{eff}_a=\alpha_a \int_{M_4}a(x^\mu)\epsilon^{abc}n^a \dot{n}^b n'^c \delta(x)\delta(y)dx^4.
\ee
Consider now a dynamic string, such that the region of vanishing $\phi_1$ is a time varying line $s^\mu(\tau,\sigma)$,  as the one described in (\ref{roto})
and (\ref{roto2}). In this situation the Dirac delta $\delta(x)\delta(y)$ is generalized to
\be\lb{gendir}
\delta(x)\delta(y)\longrightarrow \int_\Sigma \sqrt{-|\gamma|} \delta^4(x^\mu-s^\mu(\tau, \sigma)) d\tau d\sigma,
\ee
being 
\be\lb{deto}
\gamma_{\alpha\beta}=\eta_{\mu\nu}\partial_\alpha s^\mu \partial_\beta s^\nu,\qquad \alpha,\beta=\tau, \sigma,
\ee
the world sheet metric. Then, with the help of (\ref{gendir}), it is seen that effective action (\ref{ef}) is a particular case of the following general functional form
\be\lb{coup1}
S^{eff}_a=\alpha_a\int_\Sigma \int_{M_4}a(x^\mu)\epsilon^{abc}n^a \dot{n}^b n'^c \sqrt{-|\gamma|} \delta^4(x^\mu-s^\mu(\tau, z)) d\tau d\sigma dx^4,
\ee
or, after integrating in the spatial coordinates
\be\lb{coup2}
S^{eff}_a=\alpha_a\int_\Sigma a(\tau, z)\epsilon^{abc}n^a \dot{n}^b n'^c \sqrt{-|\gamma|}  d\tau d\sigma.
\ee
Here the dots correspond to derivatives with respect to $\tau$.

The formula (\ref{coup2}) gives the coupling of the axion to the orientational modes of the vortex.
In particular, for the solution (\ref{roto})
and (\ref{roto2}) found above, it is obtained that the determinant (\ref{deto}) is expressed as
\be\lb{det2}
-\gamma=(\cos^2\omega z+R^2\nu^2\sin^2\omega z)(1-R^2\nu^2)\cos^2\omega z,
\ee
after identifying $z=\sigma$ and $t=\tau$. These formulas will be employed in the next sections.

 On the other hand, a typical coupling between the axion $a$ and the scalar of the fields $\Phi^{ak}$ of (\ref{beat}) is given by
 $${\cal L}_{aq}=\lambda a^2 \text{Tr}(\overline{\Phi}_A \Phi_A-2I\Delta^2).$$
 The term $a^2$ is needed, since the axion is a pseudo Goldstone boson and an odd power will make the lagrangian pseudoscalar.  It is clear from (\ref{beat}) that this trace does not have any dependence on $U$ or, what is the same, on the orientational modes $n^i$. Thus, this coupling will solely induce 
a vertex between the axion and the translational modes. By taking into account (\ref{beat}) and the fact that $\phi_2$ is slowly varying with values near the unity while $\phi_1(r)$
tends to zero quickly near $r\sim0$, the trace can be approximated by
$$
\text{Tr}(\overline{\Phi}_A \Phi_A-2I\Delta^2)= f^2(r)\Delta^2\simeq \delta^2\Delta^2 \delta(x)\delta(y). 
$$
In the previous expression $\delta$ denotes the thickness of the string, the function $f(r)$ decays rapidly for $r>>a$ and  was approximated by the Dirac delta in the last step. 
A parametrization invariant form of the previous formula, for a non trivial loop, is the following
\be\lb{coup3}
S^{eff}_{aq}=\lambda \int a^2 \text{Tr}(\overline{\Phi}_A \Phi_A-2I\Delta^2)d^4x\simeq\lambda \Delta^2 \delta^2 \int\sqrt{-|\gamma|} a^2 \delta^{4}(x^\mu-s^\mu(\tau, z)) d\tau d\sigma d^4x.
\ee
The last expression gives the coupling between the axion and the translational modes of the vortex.

It is convenient to remark that (\ref{coup3}) give rise to a single axion emission, while (\ref{coup2}) represents two axion emission.

\section{Some formulas for axion emitted power}
The coupllings found in the previous section are fundamental for deriving the power emitted by single axion and two axion emission. For single axion emission, the main object to be calculated is transition amplitude $<S', a|S>$ from an initial string state $|S>$ to a final one $|S', a>$. This amplitude is calculated by use of LSZ formulas,  which shows as a result that
\be\lb{amp}
<S',a|S>=\int \exp(ik\cdot x)<S'|(\square+m_a^2)a(x)|S> d^4x.
\ee
In the following, the rough approximation that $|S'>\sim |S>$ will be employed, that is, the emission of a two axions does not react back on the string. This is of course a simplifying assumption since, at the end, the string excitations are expected to decay completely. The operator involved in (\ref{amp}) is calculated by means of the following formula
$$
(\square+m_a^2)a(x)=\frac{\partial L}{\partial a}.
$$
The right hand has two type of contributions, one  from the axion couplings to the translational modes and other due to the orientational ones. But the translational ones do not contribute to this process, as they involve two axion emission. The coupling to the orientational modes is obtained from (\ref{coup1}), it is simply given by
$$
\frac{\partial L}{\partial a}\bigg|_o=\alpha_a\int_\Sigma \epsilon^{abc}n^a \dot{n}^b n'^c \sqrt{-|\gamma|} \delta^4(x^\mu-s^\mu(\tau, z)) d\tau dz. 
$$
In these terms the amplitude (\ref{amp}) becomes
\be\lb{amp2}
<S',a|S>=\alpha_a\int \int_\Sigma e^{ik\cdot x}\epsilon^{abc}n^a \dot{n}^b n'^c\sqrt{-|\gamma|} \delta^4(x^\mu-s^\mu(\tau, z)) d\tau dz d^4x.
\ee
The total radiated power by axion radiation in this case can be calculated from the expression
\be\lb{poti}
P\;T=\int E|<S, a|S>|^2\frac{d^3k}{2(2\pi)^3E}.
\ee
In the last formula, $T\sim \delta(0)$ is the duration of the process, which is assumed to be infinitely large.

Consider now the amplitude corresponding to two axion emission $<S', a_1, a_2|S>$. In this case, the translational modes are the one contributing to the process.
The corresponding  amplitude is
\be\lb{amp22}
<S',a_1, a_2|S>=\int \exp(ik_1\cdot x)<S', a_2|(\square+m_a^2)a(x)|S> d^4x.
\ee
From the coupling (\ref{coup3}) of the axion to the translational modes it is found that
$$
(\square+m_a^2)a(x)=\frac{\partial L}{\partial a}\bigg|_t=2\lambda a  \text{Tr}(\overline{\Phi}_A \Phi_A-2I\Delta^2)\simeq 2\delta^2\Delta^2 \int\sqrt{-|\gamma|} a \delta^{4}(x^\mu-s^\mu(\tau, z)) d\tau dz.
$$
The searched amplitude is then
\be\lb{amp3}
<S',a_1, a_2|S>=\lambda \Delta^2\delta^2 \int \int_\Sigma e^{i(k_1+k_2)\cdot x}\sqrt{-|\gamma|} \delta^4(x^\mu-s^\mu(\tau, z)) d\tau dz d^4x.
\ee
The total radiated power in this case results
\be\lb{poti2}
P\;T=\int\int E|<S, a_1, a_2|S>|^2\frac{d^3k_1}{2(2\pi)^3k^0_1}\frac{d^3k_2}{2(2\pi)^3k^0_2},
\ee
where $E=k_1^0+k_2^0$. 

\section{Emitted power for the rotating string inspired ansatz}
It is of interest to apply the general formulas described above for the string solution (\ref{roto2}). The string location, in cartesian coordinates, is parameterized as follows
\be\lb{cartesiame}
s^t= t,\qquad s^x=\frac{\sqrt{1-R^2\nu^2}}{\omega}|\sin(\omega z)|\cos(\omega t),\qquad s^y=\frac{\sqrt{1-R^2\nu^2}}{\omega}|\sin(\omega z)|\sin(\omega t),\qquad s^z=z.
\ee
In addition, $n'^{a}=0$ for this ansatz which, together with (\ref{amp}), implies that only the translational modes contribute to the calculation of the decay.  However, this does not mean that the orientational modes are irrelevant, as they contribute to the solution given above by the parameter $R^2\nu^2$. In particular, for $R^2\nu^2\neq 0$ one has that $|\dot{s}|<1$ which means that  there are no points in the string reaching the speed of light. This is a characteristic feature for non abelian strings, and
modifies the stationary phase analysis of reference \cite{theisen}. In the following, the techniques developed in that work will be modified for dealing with the present situation.

The amplitude (\ref{amp}), applied for the solution (\ref{roto2}), is explicitly given by
$$
<S', a_1, a_2|S>\simeq 4\lambda \Delta^2\delta^2\int_{-\infty}^\infty\int_0^{\frac{\pi}{\omega}} e^{iE t} e^{-i k_x s^x-ik_y s^y-i k_z z}
$$
$$
\times \sqrt{(R^2\nu^2 \sin^2\omega z+\cos^2\omega z)(1-R^2\nu^2) \cos^2 \omega z} \;dzdt,
$$
with $E=k_1^0+k_2^0$ and $k=k_1+k_2$  the sum of the energy and the wave vector of the two axions, respectively.  By making the redefinition $k_i\to k_i/\omega$ and $E\to E/\omega$, together with the introduction of the new dimensionless variables $\eta=\omega z$ and $\xi=\omega t$, the last expression may be written as
$$
<S', a_1, a_2|S>\simeq \frac{4\lambda \Delta^2\delta^2}{\omega^2}\int_{-\infty}^\infty\int_0^{\frac{\pi}{2}} e^{-i k_x \sqrt{1-R^2\nu^2}\sin\eta\cos\xi}e^{-ik_y \sqrt{1-R^2\nu^2}\sin\eta\sin\xi}  (e^{-ik_z \eta}-e^{ik_z \eta})
$$
\be\lb{relax}
\sqrt{(R^2\nu^2 \sin^2\eta+\cos^2\eta)(1-R^2\nu^2 )\cos^2\eta}\;e^{i E \xi} d\eta d\xi,
\ee
where very elementary parity properties of the trigonometric functions were used to obtain this expression. It is important to remark that the square root, which corresponds to the world sheet metric determinant, is independent on $\xi$. This simplifies the calculation done below.

At first sight, the integral in $\eta$ may be estimated by saddle point methods and the integration in $\xi$ may be performed later on. The present author however, have found expressions that he could not handle.  For this reason, an alternative method will be employed. The use of the identity
\be\lb{ido1}
k_x \sqrt{1-R^2\nu^2}\sin\eta\cos\xi+k_y \sqrt{1-R^2\nu^2}\sin\eta\sin\xi=\sqrt{(k_x^2+k_y^2)(1-R^2\nu^2)}\sin(\eta)\sin(\xi+\delta),
\ee
$$
\sin\delta=\frac{k_x}{\sqrt{k_x^2+k_y^2}},\qquad \cos\delta=\frac{k_y}{\sqrt{k_x^2+k_y^2}},
$$
together with the integral representation of the Bessel functions of the first kind \cite{nico}
\be\lb{ido2}
J_n(\lambda)=\frac{1}{2\pi}\int_{-\pi}^\pi e^{i\lambda \sin(u)} e^{-in u}du,
\ee
yields the following Fourier expansion
\be\lb{ido3}
e^{-i k_x \sqrt{1-R^2\nu^2}\sin\eta\cos\xi-ik_y \sqrt{1-R^2\nu^2}\sin\eta\sin\xi}=\sum_{n=-\infty}^\infty e^{i n(\delta+\pi)} J_n\bigg[\sqrt{(k_x^2+k_y^2)(1-R^2\nu^2)}\sin\eta\bigg]e^{i n \xi}.
\ee
 The uniform convergence property of Fourier series imply that this equality can be integrated term by term.
Thus, by inserting the last expression into (\ref{relax}), the searched amplitude becomes a sum of the form
$$
<S', a_1, a_2 |S>\simeq  \frac{4\lambda \Delta^2\delta^2}{\omega^2}\sum_{n=-\infty}^\infty e^{i n(\delta+\pi)}\delta(E-n) a_n,
$$
with the coefficients $a_n$ given by
$$
a_n=\int_0^{\frac{\pi}{2}} (e^{-ik_z \eta}-e^{ik_z\eta}) J_n\bigg[\sqrt{(k_x^2+k_y^2)(1-R^2\nu^2)}\sin\eta\bigg] 
$$
\be\lb{an}
\sqrt{(R^2\nu^2 \sin^2\eta+\cos^2\eta)(1-R^2\nu^2) \cos^2\eta}\;d\eta.
\ee
Note that this coefficients have purely imaginary values. In addition, the property  $J_{-n}(x)=(-1)^n J_n(x)$ valid for integer $n>0$, implies that $$a_{-n}=(-1)^n a_n.$$ For the special case $R^2\nu^2=1$ the property that $J_n(0)=0$ for $n\neq 0$ implies that $a_n=0$ for $n>0$. This is desirable, as (\ref{cartesiame}) imply that there will be no motion of the coordinates of the string in this case. However, $a_0(0)$ is not necessarily zero, but it will be shown below that this term does not give any contribution to the power radiated.
In the terms given above, the squared amplitude can be rewritten as 
$$
|<S', a_1, a_2|S>|^2\simeq   \frac{32 \lambda ^2\Delta^4\delta^4T}{\omega^3}\sum_{n=0}^\infty \delta(E-n) |a_n|^2,
$$
with $T=\delta(0)/\omega$ being the duration of the process. As the axion mass is assumed to be very tiny, the momentum can be parameterized by a set of pair of polar angles
$$
k^x_i=|k^0_i| \sin \gamma_i \sin\zeta_i, \qquad k_i^y=|k^0_i|  \sin \gamma_i \cos\zeta_i,\qquad k^z_i=|k^0_i|  \cos \gamma_i,\qquad i=1,2.
$$
From here it is calculated that
$$
P=32\lambda ^2\Delta^4\delta^2\omega^2\int \int E k^0_1 dk^0_1 d\Omega_1k^0_2 dk^0_2 d\Omega_2 \sum_{n=0}^\infty \delta(E-n) |a_n|^2.
$$
Here the solid angle $d\Omega_i=\sin \gamma_i d\gamma_i d\zeta_i$ was introduced, with $i=1,2$. Note that if $R^2\nu^2$ then $a_n=0$ and $a_0\neq 0$, but the Dirac delta in the last expression forces $P=0$. This is an important consistency test, as the resulting static string should not radiate axions. By taking into account that $E=k^0_1+k_0^2$, the last  integral reduces to
\be\lb{potenciale}
P=32\lambda ^2\Delta^4\delta^4\omega^2 \sum_{n=0}^\infty \int_0^n n k^0_1(n-k^0_1) dk^0_1 d\Omega_1d\Omega_2  |a_n|^2,
\ee
where  the coefficient $a_n$ is  given by (\ref{an}) with
$$
k_x=k^0_1\sin \gamma_1 \sin\zeta_1+(n-k_1^0)\sin \gamma_2 \sin\zeta_2, \qquad k_y=k^0_1 \sin \gamma_1 \cos\zeta_1+(n-k^0_1) \sin \gamma_2 \cos\zeta_2,
$$
\be\lb{anam}
k_z=k^0_1  \cos \gamma_1+(n-k^0_1)\cos\gamma_2.
\ee
Note that, if the coefficients $a_n$ were about to be approximated by a constant value, then after integration  (\ref{potenciale}) would be a sum of terms proportional to $n^4$.
However, a careful estimation of $a_n$ should be performed, as these coefficients may go to zero and change these powers into something of the form $n^l$
with $l<4$. In addition, not only the coefficients should be estimated, but  the integral of $|a_n|^2$ over $d\Omega_1 d\Omega_2$ as well. This analysis can be done by studying limits of the Bessel functions $J_n(x)$  (\ref{an}) involved in the problem, as shown below.

For small arguments, the asymptotic behaviour of the Bessel functions of first kind is the following
\be\lb{smol}
J_n(x) \sim \frac{1}{n!}\bigg(\frac{x}{2}\bigg)^n,\qquad x<<1,\qquad n\geq 0.
\ee
This implies in particular that $J_n(0)=0$ for $n\neq 0$. 
For large arguments instead, the following behaviour holds
\be\lb{larsh}
J_n(x) \sim \sqrt{\frac{2}{x}} \cos(x-\frac{n\pi}{2}-\frac{\pi}{4})=\sqrt{\frac{1}{2x}} (e^{ix-\frac{in\pi}{2}-\frac{i\pi}{4}}+e^{-ix+\frac{in\pi}{2}+\frac{i\pi}{4}}),\qquad x>>1.
\ee
This formula is known to be valid for $n^2<x$, which is known as the Fraunhofer regime. However, the limit that will be of interest is the Fresnel limit for which $n\leq x\leq n^2$. The use of the last asymptotic formula in this regime is dubious, and may introduce some considerable error. The Fresnel regime is less studied \cite{fresnel}, but it will be described partially below. 

The coefficients $a_n$ introduced in (\ref{an}) involve expressions of the form
$$
I(E)=\int_a^b f(\gamma) e^{i E\phi(\gamma)}d\gamma, 
$$
with $f(\gamma)$ and the phase $\phi(\gamma)$ continuous and differentiable functions of the integration variable $\gamma$ and $E=|k|$. It is known that in this case, for $E\to \pm\infty$, the integral $I(E)\to 0$.
If the phase $\phi(\gamma)\neq 0$ in the interval of integration, then $I(E)$ may be approximated as in page 258 of \cite{bender}
\be\lb{bit1}
I(E)\sim \frac{ \text{sg}\phi(\gamma)}{iE}\bigg[\frac{f(b)}{\phi'(b)}e^{i E\phi(b)}-\frac{f(a)}{\phi'(a)}e^{i E\phi(a)}\bigg].
\ee
If instead there are $N$ some points $\gamma_i$ in the integration interval for which $\phi'(\gamma_i)=0$, then the saddle point approximation shows that \cite{bender}
 \be\lb{bit2}
I(E)\sim \sum_{i=1}^N e^{\frac{i\pi}{4}\text{sg}\phi''(\gamma_i)} f(\gamma_i)e^{i E\phi(\gamma_i)}\sqrt{\frac{2\pi}{E|\phi''(\gamma_i)|}}.
\ee
Assume, by use of (\ref{anam}), that roughly  $k_z\sim n$ and $\sqrt{(k_x^2+k_y^2)(1-R^2\nu^2)}\sim n\sqrt{1-R^2\nu^2}$ for $n>>1$, up to factors that depend on $k^0$, $\gamma_i$ and $\zeta_i$ and which are small only for a narrow choice of these parameters. Then, for the large region in the space described by $k^0$, $\gamma_i$ and $\zeta_i$ one has that
  $1<<|k_z|$ and  $1<< \sqrt{(k_x^2+k_y^2)(1-R^2\nu^2)}$. The Bessel functions in (\ref{an}) are in the Fresnel regime in that region. It may be assumed that they do not involve large oscillating phases. Thus, the only oscillating phase is the one involving $k_z$, which has no minima. By applying (\ref{bit1}) it follows that $a_n\sim 0$, as the determinant is zero on $\eta=\pi/2$ and the Bessel functions vanish for $\eta=0$. For $1<<|k_z|$ and  $0< \sqrt{(k_x^2+k_y^2)(1-R^2\nu^2)}<1$ the same argument holds. Thus, the radiation is concentrated in the directions $0<k_z<n_0$ with $n_0$ an integer with small or intermediate values. If $k_z$ is small and the energy $E=n$ is large, then energy conservation forces $1<<\sqrt{(k_x^2+k_y^2)(1-R^2\nu^2)}$. From (\ref{anam}) it is seen that  roughly $ \sqrt{(k_x^2+k_y^2)(1-R^2\nu^2)}\sim n \sqrt{1-R^2\nu^2}$. Thus, the expression (\ref{an}) can be approximated by
$$
a_n\sim \int_0^{\frac{\pi}{2}} (e^{-in_0 \eta}-e^{in_0\eta}) J_n\bigg(\sqrt{(k_x^2+k_y^2)(1-R^2\nu^2)}\sin\eta\bigg)
$$
\be\lb{an2}
\sqrt{(R^2\nu^2 \sin^2\eta+\cos^2\eta)(1-R^2\nu^2) \cos^2\eta}\;d\eta,
\ee
and the Bessel function in the argument may be though as in entering into the Fresnel regime. To the knowledge of the author, there is no closed expression for 
this integral. However, there exist integration formulas \cite{nico} for the Bessel function of first kind such as
\be\lb{visa1}
\int_0^{\frac{\pi}{2}}\cos(2\mu x)J_{2\nu}(2a \cos x)dx=\frac{\pi}{2}J_{\nu+\mu}(a)J_{\nu-\mu}(a).
\ee
The formula (\ref{an2}) is not exactly the same as (\ref{visa1}), the difference is due to the square root factor. However, this factor does not induce a considerable deviation from the expression (\ref{visa1}) for  $R^2\nu^2<1$. Thus, one may assume that the coefficient may be roughly approximated by
$$
a_n\sim iJ^2_n\bigg(\sqrt{(k_x^2+k_y^2)(1-R^2\nu^2)}\bigg).
$$
In brief, the discussion given above together with (\ref{anam}) suggests that main contribution to $a_n$ is concentrated in the directions defined by
$$
|k^0_1  \cos \gamma_1+(n-k^0_1)\cos\gamma_2|\leq 1, 
$$
\be\lb{contri}
1<<(k_0^2 \sin^2\gamma_1+(n-k^0)^2\sin^2\gamma_2+2 k^0(n-k^0)\sin \gamma_1\sin\gamma_2 \cos(\zeta_1-\zeta_2))(1-R^2\nu^2).
\ee
The coefficients are then 
\be\lb{hindu}
a_n\sim iJ^2_n\bigg[\sqrt{(k_0^2 \sin^2\gamma_1+(n-k^0)^2\sin^2\gamma_2+2 k^0(n-k^0)\sin \gamma_1\sin\gamma_2 \cos(\zeta_1-\zeta_2))(1-R^2\nu^2)}\bigg].
\ee
From (\ref{contri}) it is seen that main contribution comes from  angles $\cos\gamma_i<n^{-1}$, $\sin \gamma_i\sim 1$. The solid angle area subtended by these angles goes as $d\Omega_i\sim 1/n$, therefore 
$$
\int |a_n|^2 d\Omega_1 d\Omega_2\sim \frac{1}{n^2}J^4_n\bigg(n\sqrt{1-R^2\nu^2}\bigg).
$$
From this, it is concluded that the power (\ref{potenciale}) goes as 
\be\lb{potenti}
P=32\lambda ^2\Delta^4\delta^4\omega^2 \lim_{N\to \infty}\sum_{n=0}^N n^2J^4_n\bigg(n\sqrt{1-R^2\nu^2}\bigg),
\ee
The point is now to understand the behaviour for the Bessel functions in the last expression. The reference \cite{vachaspati} suggest that
$$
J'_n(a n)\sim \frac{1}{n^{\frac{2}{3}}}.
$$
This formula of course, should not necessarily be integrated in $n$ in order to find $J_n(an)$, as there may be factors of $n$ that do not correspond to the argument. In fact, integration gives a divergent result for large $n$, which is known not to be the case.
On the other hand there are recurrence formulas such as \cite{nico}
$$
J_{n+1}(x)+J_{n-1}(x)=\frac{2n}{x}J_n(x),\qquad J_{n+1}(x)-J_{n-1}(x)=2 J'_n(x).
$$
From the last formulas it may be reasonable to postulate that 
$$
J_n(a n)\sim \frac{c_n}{n^{\frac{2}{3}}}+\text{another powers}.
$$
In the following, it will be assumed that these extra powers are smaller or of almost the same order as $n^{-\frac{2}{3}}$. This is of course not a rigorous result, but by playing with large numbers in Mathematica I believe that it is a reasonable postulate. In these terms, the replacement of the sum in (\ref{potenti}) by an integral
yields the following power radiated
\be\lb{potenti2}
P=32\lambda ^2 c\Delta^4\delta^4\omega^2 N_c^{\frac{1}{3}}.
\ee
The constant $c$ arise due to the $c_n$ factors, and it has controlled values. Its specific functional form is undetermined, except that $c\to 0$ when $R^2\nu^2\to 0$.
The cutoff $N_c$ holds because this description may not be valid for energies $E>\delta^{-1}$. Its value is arguably of the order $N_c\sim \delta^{-1} \omega^{-1} \sim L\delta^{-1}$, and therefore
\be\lb{potenti3}
P=\frac{32\lambda ^2 c\Delta^4\delta^{\frac{11}{3}}}{L^{\frac{5}{3}}}.
\ee 
This result states that, for large objects, two axion radiation is suppressed. This is in qualitative agreement  with the results of \cite{davis}-\cite{theisen}, which suggest that for an extended object, with length of the order of a Parsec, the axion radiation should be subdominant with respect to gravitational radiation. However, there is no real sensibility of this result with respect to the parameter $R^2\nu^2$, except on $c$. The direction for which the power radiated by solid angle take relevant values is also not significantly deformed. Thus, it is difficult to distinguish non abelianity by studying two axion emission, at least for the solution (\ref{cartesiame}).

\section{Gravitational radiation}
The next task is to consider the gravitational radiation power emitted by the excited object. This power can be calculated with the help of formula \cite{weinberg}
\be\lb{weinberg}
P=\sum_{n=0}P_n,\qquad \frac{dP_n}{d\Omega}=\frac{G_n\omega_n^2}{\pi}[T_{\mu\nu}( \omega_n, k_n)T^{\mu\nu}(\omega_n, k_n)-\frac{1}{2}|T_{\mu}^\mu(\omega_n, k_n)|^2],
\ee
with $T_{\mu\nu}(k_n, \omega_n)$ the Fourier transform of the stress energy tensor
\be\lb{relajo}
T_{\mu\nu}(k_n, \omega_n)=\frac{1}{L}\int_0^{2L} \int_{R^3} e^{i \omega_n t-k_{ni} x^i} T_{\mu\nu}(t, x^i)d^3 x^i dt,\qquad k_{ni}k_n^i=\omega_n^2
\ee
The frequencies $\omega_n= n \pi/L$. There are alternative formulas such as the quadrupole approximation, but they have additional assumptions such as that the emitted wavelength by the source is larger than its  size. Instead, the formula (\ref{weinberg}) employs a smaller amount of hypothesis, and for this reason it is the one to be applied here.

The formula (\ref{weinberg}) shows that the stress energy tensor $T_{\mu\nu}(x^i, t)$ is the main quantity to be found. It can be obtained by varying the action (\ref{manta}) of the string, which it is written again below by further reference
\be\lb{manta2}
S=T \int \sqrt{-|h|}h^{ab}g_{\mu\nu}\partial_a s^\mu \partial_b s^\nu \sqrt{-\gamma} \delta^4(x^\mu-s^\mu(\sigma, \tau))d^4x.
\ee
In the conformal gauge, this action is
$$
S=T\int\eta^{ab}\bigg[ \widetilde{g}_{\mu\nu}\partial_a s^\mu \partial_b s^\nu+ R^2 \partial_a \alpha \partial_b\alpha
+R^2\sin^2\alpha \partial_a \beta \partial_b \beta\bigg] \sqrt{-\gamma} \delta^4(x^\mu-s^\mu(\sigma, \tau)) d^4 x.
$$
In the last expression $\eta^{ab}=(-1, 1)$, and the change of notation $g_{\mu\nu}\to \widetilde{g}_{\mu\nu}$ for the translational modes target metric has been made.
This change of notation is convenient, since the determinant
$$
\sqrt{-\gamma}=\sqrt{-g_{\mu\nu}\partial_a s^\mu \partial_b s^\nu},
$$
is the only quantity in the lagrangian related to the space time metric $g_{\mu\nu}$. In the present situation, $g_{\mu\nu}$ and $\widetilde{g}_{\mu\nu}$ coincide with the Minkowski metric $\eta_{\mu\nu}$. However, they have not to be identified, as one is representing a sigma model arising from details of vortex interactions and the other is representing the space time geometry the vortex is embedded in. In addition, it is $g_{\mu\nu}$ the one that has to be considered for calculating the stress energy tensor $T_{\mu\nu}$.
This tensor, in the conformal gauge, is given by
$$
T^{\mu\nu}=T\eta^{ab}\bigg[ \widetilde{g}_{\gamma\delta}\partial_a s^\gamma \partial_b s^\delta+ R^2 \partial_a \alpha \partial_b\alpha
+R^2\sin^2\alpha \partial_a \beta \partial_b \beta\bigg]\frac{1}{\sqrt{-\gamma}}\frac{\delta \gamma}{\delta g_{\mu\nu}}\delta^4(x^\mu-s^\mu(\sigma, \tau)).
$$
By use of the Jacobi formula for the determinant, it follows that
$$
\frac{\delta \gamma}{\delta g_{\mu\nu}}=\gamma \gamma^{ab} \frac{\delta \gamma_{ab}}{\delta g_{\mu\nu}}=\gamma \gamma^{ab}  \partial_a s^\mu \partial_b s^\nu.
$$
By taking into account the last three formulas it is obtained the following expression for the stress energy tensor of the configuration
$$
T^{\mu\nu}=-T \eta^{ab}\bigg[ \widetilde{g}_{\gamma\delta}\partial_a s^\gamma \partial_b s^\delta+ R^2 \partial_a \alpha \partial_b\alpha
+R^2\sin^2\alpha \partial_a \beta \partial_b \beta\bigg]
\gamma^{cd}  \partial_c s^\mu \partial_d s^\nu  \sqrt{-\gamma}\delta^4(x^\mu-s^\mu(\sigma, \tau)).
$$
By use of the last expression and (\ref{cartesiame}) it is calculated that
$$
T^{tt}=-\frac{2T(1-R^2\nu^2)\cos^2\omega z}{\cos^2\omega z+R^2\nu^2\sin^2\omega z}\sqrt{-\gamma}\delta^4(x^\mu-s^\mu(\sigma, \tau)),
$$
$$
T^{tx}=2T(1-R^2\nu^2)\cos^2\omega z\frac{\sqrt{1-R^2\nu^2}|\sin\omega z| \sin\omega t}{\cos^2\omega z+R^2\nu^2\sin^2\omega z}
\sqrt{-\gamma}\delta^4(x^\mu-s^\mu(\sigma, \tau)),
$$
$$
T^{ty}=-2T(1-R^2\nu^2)\cos^2\omega z\frac{\sqrt{1-R^2\nu^2}|\sin\omega z| \cos\omega t}{\cos^2\omega z+R^2\nu^2\sin^2\omega z}
\sqrt{-\gamma}\delta^4(x^\mu-s^\mu(\sigma, \tau)),
$$
$$
T^{xx}=-2T(1-R^2\nu^2)\cos^2\omega z\bigg[\cos^2\omega t
-\frac{(1-R^2\nu^2)\sin^2\omega z \sin^2\omega t}{\cos^2\omega z+R^2\nu^2\sin^2\omega z}\bigg]\sqrt{-\gamma}\delta^4(x^\mu-s^\mu(\sigma, \tau)),
$$
$$
T^{yy}=-2T(1-R^2\nu^2)\cos^2\omega z\bigg[\sin^2\omega t-\frac{(1-R^2\nu^2)\sin^2\omega z \cos^2\omega t}{\cos^2\omega z+R^2\nu^2\sin^2\omega z}
 \bigg]\sqrt{-\gamma}\delta^4(x^\mu-s^\mu(\sigma, \tau)),
$$
\be\lb{golubo}
T^{xy}=-2T(1-R^2\nu^2)\cos^2\omega z\sin \omega t \cos\omega t\bigg[\frac{(1-R^2\nu^2)\sin^2\omega z}{\cos^2\omega z+R^2\nu^2\sin^2\omega z}
 +1\bigg]\sqrt{-\gamma}\delta^4(x^\mu-s^\mu(\sigma, \tau)).
\ee
From this point, the procedure of calculating the power radiated is quite analogous to the axion case. The energy momentum tensor is of the form 
$$
T_{\mu\nu}(x^i, t)=T A_{\mu\nu}(x^i, t)\sqrt{-\gamma}\delta^4(x^\mu-s^\mu(\sigma, \tau)).
$$
The formula (\ref{relajo}) is the analogous of (\ref{relax}), but the square of the metric determinant $\sqrt{-\gamma}$ is replaced by  $A_{\mu\nu}(x^i, t)\sqrt{-\gamma}$.  In the same fashion than for the square of the world sheet metric determinant $\sqrt{-\gamma}$, the factors  $A_{\mu\nu}(\omega z, \omega t)$ are simple periodic functions whose values are not far from unity. The Fourier transform of $T_{\mu\nu}$ is then 
$$
T_{\mu\nu}(k_n, \omega_n)=\frac{T}{L\omega^2} \int_{0}^{2\pi}\int_0^{\pi} e^{-i k_x \sqrt{1-R^2\nu^2}\sin\eta\cos\xi}e^{-ik_y \sqrt{1-R^2\nu^2}\sin\eta\sin\xi}  e^{-ik_z \eta}
$$
\be\lb{furia}
A_{\mu\nu}(\eta,\xi)\sqrt{(R^2\nu^2 \sin^2\eta+\cos^2\eta)(1-R^2\nu^2 )\cos^2\eta}\;e^{i \omega_n \xi} d\eta d\xi,
\ee
where the replacement $E\to E/\omega$ and $k\to k/\omega$ has been performed. The quantities $A_{\mu\nu}(\eta,\xi)$ are functions the two variable integrations. At first sight, this dependence makes  the  steepest descent method employed in previous sections more difficult to apply. However, from (\ref{golubo}) it is seen that the dependence in $\xi$ is very simple. It is given by  linear combinations of the trigonometric functions $\cos\xi$, $2\cos^2\xi=1+\cos(2\xi)$, $\sin \xi$, $2\sin^2 \xi=1-\cos(2\xi)$ and $2\sin\xi \cos\xi=\sin(2\xi)$. This makes the calculation of $T_{\mu\nu}(k_n, \omega_n)$ much easier than expected.

Consider for example the Fourier component $T_{xt}(k_n,\omega_n)$. From (\ref{golubo}) and (\ref{furia}) it is seen that
$$
T_{xt}(k_n, \omega_n)=\frac{iT(1-R^2\nu^2)^{\frac{3}{2}}}{L\omega^2} \int_{0}^{2\pi}\int_0^{\pi} e^{-i k_x \sqrt{1-R^2\nu^2}\sin\eta\cos\xi}e^{-ik_y \sqrt{1-R^2\nu^2}\sin\eta\sin\xi}  e^{-ik_z \eta}
$$
$$
\frac{|\sin\eta| |\cos\eta|^3}{\sqrt{\cos^2\eta+R^2\nu^2\sin^2\eta}} (e^{i (n+1) \xi}-e^{i(n-1)\xi})\;d\eta d\xi.
$$
By employing now (\ref{ido1})-(\ref{ido3}) and by assuming that the integration order may be changed, the following Fourier components are obtained
$$
T_{xt}(k_n, \omega_n)=\frac{iT(1-R^2\nu^2)^{\frac{3}{2}}e^{in\delta}}{L\omega^2}\int_0^{\pi}  \frac{e^{-ik_z \eta}|\sin\eta| |\cos\eta|^3}{\sqrt{\cos^2\eta+R^2\nu^2\sin^2\eta}}
$$
$$
\bigg[e^{-i\delta} J_{n+1}\bigg(\sqrt{(k_x^2+k_y^2)(1-R^2\nu^2)}\sin\eta\bigg)-e^{i\delta} J_{n-1}\bigg(\sqrt{(k_x^2+k_y^2)(1-R^2\nu^2)}\sin\eta\bigg)\bigg]d\eta.
$$
This quantity can be estimated in terms of (\ref{bit1})-(\ref{bit2}) by making a direct analogy with (\ref{an}).  This analogy follows by replacing 
$$
\sqrt{(R^2\nu^2 \sin^2\eta+\cos^2\eta)(1-R^2\nu^2) \cos^2\eta}\longrightarrow \frac{|\sin\eta| |\cos\eta|^3}{\sqrt{\cos^2\eta+R^2\nu^2\sin^2\eta}},
$$
in (\ref{an}) and by taking into account that now there are two Bessel functions involved, with label $n+1$ and $n-1$ instead of $n$. In these terms, formulas analogous to the ones obtained in the previous section for $a_n$ can be found for $T_{xt}(k_n, \omega_n)$. The same type of formulas can be found for the other components of $T_{\mu\nu}(k_n, \omega_n)$ as well. However, there is no need to go through all this calculation in order to estimate the power emitted. By simply parameterizing the momentum as
$$
k_x=n \sin \gamma \sin\zeta, \qquad k_y=n \sin \gamma \cos\zeta,\qquad
k_z=n  \cos \gamma,
$$
it follows that the least decaying contributions to $T_{\mu\nu}(k_n, \omega_n)$ are, as before, proportional to $n^{-\frac{2}{3}}$. Thus, the power radiated formula (\ref{weinberg}) then may be estimated as \be\lb{daleduro3}
P\sim \frac{G_n cT^2}{\pi L^2\omega^2} (1-R^2\nu^2)^2\sum_{n=n_0}^\infty \frac{c_n}{n^{\frac{5}{3}}},
\ee
up to the sum of the first terms, for which the steepest descent method for calculation does not apply, but which have moderate values as well.
The sum in the last expression is  convergent, thus
\be\lb{daleduro3}
P\sim G_n cT^2(1-R^2\nu^2)^2,
\ee
with $c$ an undetermined constant, which may depend on the internal parameter $R^2\nu^2$. 

For abelian strings, this radiation was studied in several references \cite{galaxy}-\cite{kibble}. In particular, the authors of \cite{vachaspati}-\cite{burden} postulate, for closed loops, that
\be\lb{vacha}
P=\gamma_l G_n T^2,
\ee
where $\gamma_l$ is a factor which depends on the shape of the loop but is independent on its perimeter $L$. Both (\ref{vacha}) and (\ref{daleduro3}) have the same dependence. At first sight, formula (\ref{daleduro3}) does not exhibit a sensible dependence on the internal rotations, unless $R^2\nu^2\to 1$. But in fact, this dependence may be important for detecting non abelianity. The point is that, in a Gedanken experiment, a loop with a definite shape and tension $T$ emits  with a power of the form (\ref{vacha}). However, if the factor $\gamma_l$ describes a loop with the wrong shape, it may be an indication that the vortex is in fact non abelian.

 \section{Internal rotations}
In the previous section, it was found that for the rotating string inspired ansatz (\ref{roto2})  two axion emission power is not very sensitive to the excitations of the internal motion. The gravitational emission instead may give a hint about non abelianity due to a deviation from the shape factor $\gamma_l$.  However, single axion emission (\ref{amp2})-(\ref{poti})
was not yet considered, as (\ref{roto2}) does not allow this process, at least at perturbative level. It may be of interest to find situations in which this emission is present.
In fact, it is not the quantitative form of the power radiated for single axion emssion which points out non abelianity. Instead, it is the fact that single axion emission is present that signals it, as single axion emission (\ref{amp2})-(\ref{poti}) is only possible when orientational modes are present. The point is to determine if this effect is of the order or larger than the previously discussed, or if it is suppressed instead.

In the following, inspired by the spike string solution of \cite{kruczenski}, it will be assumed that the string is straight and located at $r=0$. The perturbations to be considered below are purely internal, that is, the string position is always the same straight line $r=0$.
The perturbation is such that only the sphere unit vector $n^a$ is prompted to a varying function of $z$ and time.
The anzatz to be use for this situation is \cite{kruczenski}
\be\lb{onza}
t=R \tau,\qquad \alpha=\alpha(\sigma),\qquad \beta=\omega \tau+\sigma,\qquad z=\lambda \sigma.
\ee
Here $2\pi\lambda=L$, with $L$ the length of the vortex, and is not to be confused with the coupling present in the interaction term (\ref{coup3}). The parameter $\sigma$ will be chosen proportional to the coordinate $z$ of the string, and $\tau$ is a time coordinate. Both $\tau$ and $\sigma$ are dimensionless. Also $\omega$ is dimensionless, and is not to be confused with the frequency $\omega=\pi/L$ of the previous sections.

It should be recalled that (\ref{onza}) is not a true solution of the system composed by the equations derived from (\ref{logrado}) together with the conformal constraints (\ref{constro})-(\ref{constro2}). The equations of motion are in fact satisfied but the conformal constraint (\ref{constro2}) is not\footnote{It is a true solution for the reference \cite{kruczenski} however, but these authors are studying  strings in other curved backgrounds.}. This is expected from the physical point of view and is indicating that an straight vortex
with varying orientational moduli tend to move. The explicit form of the true solution may be in fact complicated. In the following this vortex movement will be neglected and we conform ourselves to use (\ref{onza}), as we are intended to study the axion emission due to pure orientational modes.  

The position of the string is static, but its moduli is varying with the height and time. The only unknown function in the previous anzatz is $\alpha(\sigma)$, and the solution to be employed is \cite{kruczenski}
\be\lb{kru}
\alpha'=\frac{\sin\alpha}{C}\sqrt{\frac{\omega^2\sin^2\alpha-C^2}{1-\omega^2 \sin^2\alpha}}.
\ee
There are several cases to consider.  In the situation $\omega^2<1$ it is clear that $C^2<\omega^2$ since otherwise the square root would be imaginary. Thus $C^2<\omega^2<1$.
The values of the angle are such that 
$$
\frac{C^2}{\omega^2}<\sin^2\alpha<1.
$$
 In the other case $\omega^2>1$ one has three situations. These are $C^2<1<\omega^2$, for which 
 $$
\frac{C^2}{\omega^2}<\sin^2\alpha<\frac{1}{\omega^2},
$$
 $1<C^2<\omega^2$ for which
 $$
\frac{1}{\omega^2}<\sin^2\alpha<1,
$$
 or $1<\omega^2<C^2$, which leads to the bound
  $$
\frac{1}{\omega^2}<\sin^2\alpha<\frac{C^2}{\omega^2}.
$$
The rotating vectors $n^a$ satisfying that $n^a n^a=1$, and which define the internal orientation of the string, are given by
$$
n^1=\sin \alpha \cos(\omega \tau+\sigma),\qquad n^2=\sin \alpha \sin(\omega \tau+\sigma),\qquad n^3=\cos\alpha.
$$
In these terms it follows that
$$
\epsilon^{abc}n^a\dot{n}^b n'^c=-\omega\alpha'\sin \alpha-\omega \cos\alpha\sin^2\alpha \sin(\omega\tau+\sigma)\cos(\omega\tau+\sigma).
$$
From the last expression is calculated that
$$
<S,a|S>=-\alpha_a \int_{-\infty}^\infty e^{iE R \tau}\bigg(\int_0^{2\pi}e^{-ik_z\lambda \sigma} [\omega\alpha'\sin \alpha+\frac{\omega}{2}  \cos\alpha\sin^2\alpha \sin(2\omega\tau+2\sigma)]d\sigma\bigg) d\tau
$$
By further parameterizing $k_z=E\cos\gamma$ it follows that
$$
<S,a|S>=\alpha_a I_1\delta(ER)+i\alpha_a\delta(ER-2\omega)I_2-i\alpha_a\delta(ER+2\omega)I_2^\ast.
$$
with
\be\lb{jarmoni}
I_1=\int_0^{2\pi}e^{-iE\cos\gamma\lambda \sigma}\omega\alpha'\sin \alpha d\sigma,\qquad I_2=\int_0^{2\pi} \frac{\omega}{2}  \cos\alpha\sin^2\alpha e^{-2i\sigma-i E\cos\gamma\lambda\sigma}d\sigma.
\ee
Then
$$
|<S,a|S>|^2=\frac{\alpha_a^2}{R^2} I^2_1\delta(E) \delta(0)+\frac{\alpha_a^2}{R^2} |I_2|^2 [\delta(E-\frac{2\omega}{R})+\delta(E+\frac{2\omega}{R})]\delta(0),
$$
and the power radiated 
$$
P=\frac{dE}{dt}T=\frac{1}{(2\pi)^3}\int \frac{d^3k}{2E} E|<S,a|S>|^2,
$$
with $T=2\pi \delta(0)$ is given explicitly by
$$
P=\frac{\alpha_a^2}{R^2}\frac{1}{(2\pi)^3}\int_0^\infty\int_0^{2\pi}\int_0^\pi \frac{E^2 dE \sin \gamma d\gamma d\zeta}{2} [I^2_1\delta(E)+ |I_2|^2 [\delta(E-\frac{2\omega}{R})+\delta(E+\frac{2\omega}{R})].
$$
From the second (\ref{jarmoni}) it follows that 
$$
P=\frac{\alpha_a^2\omega^2}{(2\pi)^3R^4}\int_0^{2\pi}\int_0^\pi  |I_2|^2\sin \gamma d\gamma d\zeta,
$$
where now 
$$
 I_2=\int_0^{2\pi} \frac{\omega}{2}  \cos\alpha\sin^2\alpha e^{-2i(1- \frac{\omega\lambda}{R}\cos\gamma)\sigma}d\sigma.
$$
If the parameter $\lambda$ is large in comparison with $R$ and $\cos\gamma>R (\omega \lambda)^{-1}$, then the formula (\ref{bit1}) can be employed for estimating $I_2$.
The result is of the form
$$
I_2\sim \frac{R}{\omega \lambda}.
$$
The same follows for  $\cos\gamma<R (\omega \lambda)^{-1}$, as the interval of integration has the small length $l=R (\omega \lambda)^{-1}$.
In these terms, the power is estimated as
\be\lb{penle}
P\sim\frac{\alpha_a^2}{R^2L^2},
\ee
up to numerical factors with controlled values. The  power emitted seems to decay for large vortices more rapidly than for two axion emission, but of course this result is not rigorous as the employed solution is just an approximation. 

\section{Discussion}
In the present work it was argued that the main consequence of non abelianity, for certain type of $SU(2)$ gauge theories, is the modification of gravitational the loop factor $\gamma_l$ in (\ref{vacha}). Another difference is that the existence of single axion emission process. The task is now to understand in which regimes one or other process is dominant. Consider a phase similar to the CFL phase. It may be assumed that the length $\alpha_a$ in (\ref{penle}) and the thickness of the string $\delta$ are of the form
$$
\alpha_a=\frac{1}{\sqrt{T}} f\bigg(\frac{\mu}{\Delta}\bigg),\qquad \delta=\frac{1}{\sqrt{T}} g\bigg(\frac{\mu}{\Delta}\bigg),
$$
with $f(x)$ and $g(x)$ unknown functions which takes moderate values for $\mu \sim \Delta\sim T_c$. By use of the formula
 (\ref{tensito}) for the tension and  (\ref{erre2}) for the radius $R$, it follows that the one axion emission given for (\ref{penle}) is predominant over (\ref{potenti3}) for a size $L$ given by
 $$
 L<\frac{\Delta^5}{\mu^6}\bigg(\frac{\mu}{T_c}\bigg)^{11}  h\bigg(\frac{\mu}{\Delta}\bigg),
 $$
 with $h(x)$ also an unknown function.
 For values of the chemical of $\mu\sim 10^2-10^3$ MeV, $T_c\sim 100$ MeV and $\Delta \sim 300$ MeV, under the assumption that $h(x)$ takes moderate values,  the resulting length is very small $10^{-13}$ m$<L<10^{-4}$ m.  Thus, it is likely that one axion emission is a suppressed process, except for microscopically small objets, although the calculation presented here has not been rigourous. On the other hand, two axion emission (\ref{potenti3}) is predominant over gravitation (\ref{daleduro3}) for 
 $$
 L<\bigg(\frac{T_c}{\mu}\bigg)^{\frac{23}{5}}\bigg(\frac{M_p^{6}}{\Delta^{11}}\bigg)^{\frac{1}{5}} k\bigg(\frac{\mu}{\Delta}\bigg),
 $$
 with $k(x)$ an unknown function with moderate values when the parameters take the values  just mentioned. A typical limit value is $L\sim 10^{9}$ m, which is $10^{-2} d_{es}$ with $d_{es}$
 the distance between the earth and the sun. For an object of such size, or even larger, gravitational effects start to predominate. For a parsec scale, gravitational radiation dominates completely over axion radiation. This is of course for energy scales characteristic of the CFL phase, for other models this calculation has to be repeated, but gravitational radiation will predominate at a very large scale. 
 
 There may the case in which there is no coupling between the axion and the translational modes. In this case the typical length for which
 gravity dominates over single axion emission is $L\geq 100$ km.

 The picture above change for very small values of the symmetry breaking scale $\Delta$. For $\Delta \sim m_a\sim 10^{-5}$ eV, single axion emission dominates over two axion emission until  $L\sim 10$ m. For scales typical of dark energy model, of the order  $\Delta\sim 10^{-28}$ eV, single axion emission dominates until $L\sim 10^{24}$ m, a scale much larger than a parsec. A priori, detection of no abelianity is simpler in this case, but it is more difficult to access to the physics at these small energies and large scales.
 
 The discussion given above is suggesting that  single axion emission due to orientational modes may be irrelevant, and the importance of these modes arise by modifying the gravitational loop factor $\gamma_l$. However, there may be other situations, not considered here due to technical complications, for which this picture may change. The internal space considered here is $S^2$, but there are $SU(N)\times U(1)$ gauge theories whose internal space is $CP(N-1)$. Perhaps the presence of a larger amount of directions allow to find solutions for which the vortex position is not of the form $s^i=a^i(x+t)+b^i(x-t)$, which may modify the analysis made in \cite{vachaspati}-\cite{burden}. It is likely that for these hypothetical solutions (\ref{vacha}) still holds, but the correction of the loop factor may be more pronounced. Another interesting line of work would be to consider semi-local strings \cite{novedad1}-\cite{novedad10}, which have internal orientation space which is not compact. The presence of non compact directions can generate a much more rich space of solutions for moving strings. It may be of interest to study these solutions and how the vortex position evolves for these models, together with axion and gravitational emission. This of course technically more complicated, as the couplings between axion and the string and the  equations of motion in this case are more involved. The study of emission channels for these largely non abelian objects deserves, in my opinion, further attention.  
\section*{Acknowledgments}
O. S is supported by CONICET, Argentina.

\end{document}